\documentclass[useAMS,usenatbib]{mn2e}

\newcommand{\msun}{M$_{\sun}$}
\newcommand{\kms}{km s$^{-1}$}
\newcommand{\msuns}{M$_{\sun}~$}
\newcommand{\kmss}{km s$^{-1}~$}

\newcommand{\mnras}{MNRAS}
\newcommand{\apj}{ApJ}
\newcommand{\apjl}{ApJ}
\newcommand{\aj}{AJ}
\newcommand{\aap}{A\&A}

\newcommand{\apss}{APSS}
\newcommand{\apjs}{ApJS}

\newcommand{\pasj}{PASJ}
\newcommand{\pasp}{PASP}
\newcommand{\planss}{P\&SS}
\newcommand{\na}{New Ast.}

\newcommand{\nbody}{{\it N}-body~}

\usepackage{graphicx,amssymb,amsmath}

\title[Primordial triples and collisions of massive stars]
{Primordial triples and collisions of massive stars}
\author[Moeckel \& Bonnell]{Nickolas Moeckel$^{1}$\thanks{{\tt nickolas1@gmail.com; @nickolas1}} and Ian A. Bonnell$^{2}\thanks{{\tt iab1@st-andrews.ac.uk}}$\\
$^{1}$University Observatory Munich, Scheinerstrasse 1, D-81679 Munich, Germany\\
$^{2}$SUPA, School of Physics and Astronomy, University of St Andrews, North Haugh, St Andrews, Fife, KY16 9SS}

\begin{document}

\date{Accepted XXX. Received XXX; in original form XXX}

\pagerange{\pageref{firstpage}--\pageref{lastpage}} \pubyear{XXXX} 

\maketitle

\label{firstpage}

\begin{abstract}
Massive stars are known to have a high multiplicity, with examples of higher order multiples among the nearest and best studied objects. In this paper we study hierarchical multiple systems (an inner binary as a component of a wider binary) of massive stars in a clustered environment, in which a system with a size of 100--1000 au will undergo many close encounters during the short lifetime of a massive star. Using two types of \nbody experiment we determine the post-formation collision probabilities of these massive hierarchies. We find that, depending on the specifics of the environment, the hierarchy, and the amount of time that is allowed to pass, tens of percent of hierarchies will experience a collision, typically between the two stars of the inner binary. In addition to collisions, clusters hosting a hierarchical massive system produce high velocity runaways at an enhanced rate. The primordial multiplicity specifics of massive stars appear to play a key role in the generation of these relatively small number events in cluster simulations, complicating their use as diagnostics of a cluster's history.
\end{abstract}

\begin{keywords}
binaries: general -- methods: n-body simulations -- open clusters and associations: general  -- stellar dynamics
\end{keywords}

\section{Introduction}
\label{introduction}

Collisions between stars is a topic that, depending on its context, can carry with it a whiff of controversy. Several authors have studied it as a mechanism for general massive star formation \citep[e.g.][]{bonnell98a, bally05}, although the prime motivation for considering collisional formation has weakened as numerical models of star formation have become more sophisticated. More recent work on collisions in the formation phase of massive stars has tended to focus on the most massive, exotic objects \citep{davis10,moeckel11, baumgardt11}. This scenario bears many similarities to the runway collision product route to intermediate mass black holes in core collapsed massive clusters \citep[e.g.][]{portegies-zwart02, gurkan04, freitag06,gaburov10}.

This latter scenario has its own controversies, but collisions of stars themselves in the purely gravitational \nbody studies devoted to it is not one of them. Given sufficiently high stellar densities and enough time, collisions will occur. As another example, collisions have been invoked in order to explain blue stragglers \citep[e.g.][]{leonard89,sigurdsson93,bacon96,fregeau04}. While most of these studies are purely gravitational and involve only binaries and singles, recent work has pushed into higher order multiples \citep{leigh12}. Beyond gravity, other work examines the hydrodynamics of stellar collisions \citep[e.g.][]{davies93,sills97,sills02,lombardi03,laycock05,gaburov08a,gaburov10}.

Our focus here is on collisions involving massive stellar systems (with masses greater than 10 \msun) after their formation and during their brief main sequence lifetime, when they are gravitationally interacting with other stars in normal\footnote{By which we mean non-extreme, e.g. without invoking temporarily high stellar densities at some stage of a star cluster's early existence.} environments. Our main motivation is the observation that massive stars are highly multiple \citep{garcia01,sana08, sana09,chini12,sana12}; we lay the scenario we have in mind in more detail now.

\subsection{Encounter timescales}
Consider first encounters between a binary system and a passing star with a maximum closest approach between the intruder and the binary centre of mass $r_{enc}$. For something interesting to happen, $r_{enc}$ should be something like the binary semi-major axis $a$. We can estimate the frequency of these encounters as $t_{enc} \approx (n \sigma v)^{-1}$, with the number density of single stars $n$ moving at velocity $v$ relative to the binary with an encounter cross section $\sigma$. The cross section must take into account the gravitational focussing of orbits \citep[e.g.][]{leonard89}, so that for a binary of mass $M_b$ and an intruder of mass $M_i$ the cross section is 
\begin{equation}
	\sigma = \pi r_{enc}^2 \left(1 + \frac{2 G (M_b + M_i)}{r_{enc}v^2} \right).
\end{equation}

\begin{figure}
 \includegraphics[width=80mm]{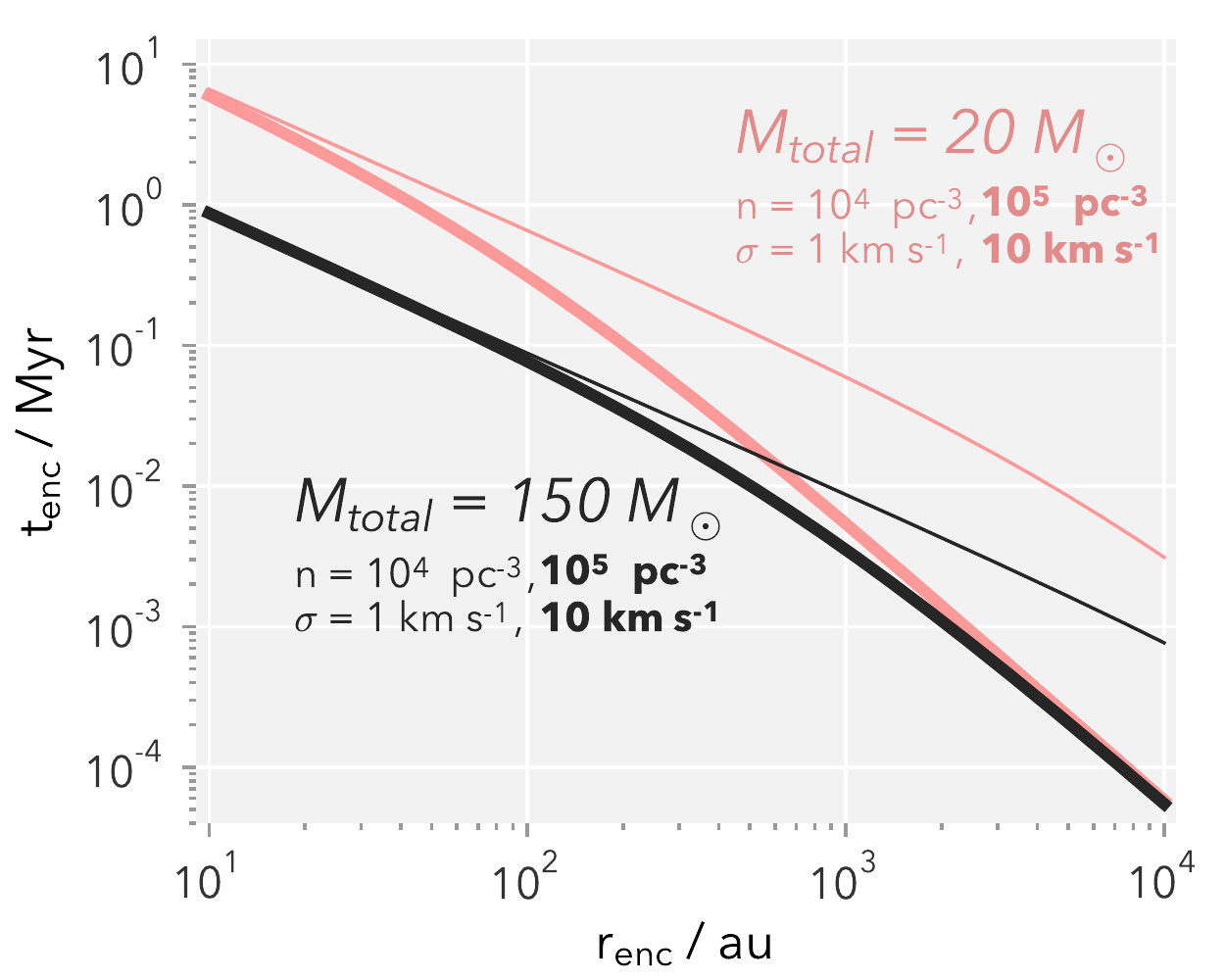}
 \caption{The encounter timescale versus encounter radius for representative massive stellar systems. The red line shows an encounter between 20 \msuns of stars, the black lines for 150 \msun. Thin lines are in an environment with stellar number density $n = 10^4$ pc$^{-3}$ and one dimensional velocity dispersion $\sigma = 1$ \kms, and the thick lines are for $n = 10^5$ pc$^{-3}$ and $\sigma = 10$ \kms.}
  \label{timescales}
\end{figure}

In figure \ref{timescales} we show this encounter timescale\footnote{We actually plot the timescale averaged over a Maxwellian velocity distribution \citep{binney08}, which introduces a correction of order unity to the $n\sigma v$ estimate.} as a function of $r_{enc}$ for representative values of the system mass and cluster environment. At small encounter separations the gravitational focussing limit depends on the stellar masses; the large separation geometrical limit is set by the cluster environment. Even very massive systems (encounter partners totalling over 100 \msun) have encounter timescales of order 1 Myr for encounter radii of 10 au. At 100s to 1000s of au, any massive system in a typical cluster has an encounter timescale of a fraction of a Myr. Binaries in this separation range should have close encounters during the stars' main sequence lifetimes.

\subsection{The insufficiency of mere binaries}
Encounter frequency is not enough to cause collisions, however; the multiple systems should also be resistant to quick and outright destruction by the intruder and, furthermore, likely to have individual stars pass closely enough together during the interaction to collide. A binary with component masses $m_0$ and $m_1$ and semi-major axis $a$ that is energetically `soft' relative to the energy of a typical intruder star of mass $M_i$ with velocity dispersion $\sigma$, i.e.
\begin{equation}
	\left| \frac{G m_0 m_1}{2 a} \right | \lesssim \frac{1}{2}M_i \sigma^2,
\label{hardsoft}
\end{equation}
will be quickly disrupted without undergoing any complex small-{\it N} dynamics \citep{heggie75, hut83}. A `hard' binary by contrast, with binding energy well in excess of the typical kinetic energy of it's neighbours, is more likely to undergo chaotic and complex resonant encounters. Massive binaries are energetically hard out to large enough separations to allow interesting encounters to take place on timescales small compared to the short lifetimes of O stars.

However, with the wider separation relative to the stellar radius comes a decreasing chance of collisions \citep{fregeau04}, and the increased encounter rate does not generate a higher net collision frequency. Generally, for an encounter to have a high probability of collisions between stars you need binaries that are so compact ($R_{\star} / a \lesssim 10^{-3}$, or $a \lesssim 10$ au for massive stars) that in a cluster with peak number densities of order $10^4$ pc$^{-3}$ the encounter rate is too small to be significant over a few Myr.

What if one of the stars in a binary that is large enough to have frequent encounters is not a single, but itself a compact binary? Generally referred to as a hierarchical system, this setup introduces the possibility of a two-stage collision process: the wide (but still energetically hard) binary acts as a net to draw passing stars into encounters at a high rate. The interaction with the wide binary shepherds intruders into close proximity with the compact binary with a higher frequency than it would experience by itself, potentially leading to collisions. \citet{leigh11} analytically studied this effective enhancement of the inner binary interaction rate in the context of blue straggler formation; \citet{geller13} numerically confirmed the predictions of that work. The process is relevant to dynamically formed triples given the time and length scales of old open and globular clusters, and here we investigate it in young clusters with massive primordial triples.

\subsection{Observational and theoretical support for higher order massive multiples}
Higher order multiples are found among nearby stars of approximately Solar mass \citep[e.g.][]{duquennoy91,raghavan10}, as well as the nearby massive stars in the Trapezium \citep[$\Theta^1$ Ori A and B are a triple and at least quadruple;][]{preibisch99, schertl03} and elsewhere in Orion; $\Theta^2$ Ori A \citep{preibisch01} and $\sigma$ Ori AB \citep{sanz-forcada04}. In the embedded phase there are already hints of high order multiplicity among massive objects \citep[e.g. W3 IRS 5;][]{megeath05}, suggesting a primordial rather than than dynamic origin.

On the theoretical side, perhaps the most likely formation scenario for primordial massive binaries is disc fragmentation \citep{adams89,laughlin94,bonnell94,bonnell94a,bonnell94b}. In simulations of star formation, starting from a variety of initial conditions and utilising different techniques, small-{\it N} multiples born out of disc fragmentation are common \citep[e.g.][]{krumholz07a, krumholz09, bate09b, peters10, stamatellos11,greif11,bate12}. Around massive stars in particular fragmentation seems prevalent, perhaps because in contrast to Solar type stars continued high accretion rates onto the disc result in conditions more favourable to fragmentation into binaries of multiple systems \citep{kratter08,kratter10}. Fragmentation in such massive discs tends to occur at the disc edge, at radii of order $10^2$ to $10^3$ au, which is the right separation for frequent encounters in a cluster. 

A possible scenario for the formation of a hierarchical triple is fragmentation followed by accretion-induced orbital shrinkage \citep{bonnell05} or disc migration \citep{goldreich80} leading to a binary with separation of order 10 au, surrounded by a circumbinary disc \citep{artymowicz94, artymowicz96}. Further fragmentation of this disc could lead to another companion at larger radii. Alternatively, early protostellar dynamics involving a disc that fragments into multiple objects may settle into a hierarchical arrangement \citep[e.g.][]{sterzik98}, particularly when dissipative tidal forces are included \citep{mardling01}. 

While hydrodynamic simulations of star formation create higher order multiples with some frequency, they are seldom included in initial conditions when performing gravitational \nbody studies of star clusters (although they certainly can form during the course of a simulation). Encounters involving triples have only recently seen some of the attention given to binary-single and binary-binary encounters \citep{leigh12}, and their focus was on stars of roughly Solar mass. If massive hierarchies are included in the initial conditions, the timescale and energetics arguments above are at least suggestive of resultant interesting encounters.

\subsection{The plan}
In this paper we set out to numerically determine stellar collision rates involving massive, primordial, hierarchical triples in a clustered environment. We make use of two types of numerical experiment. In section \ref{smallnsetup} we describe idealised encounters between isolated hierarchies and perturbing stars, and in section \ref{smallnresults} we describe those results, and in section
\ref{coplanarscattering} we briefly discuss their restriction to coplanar hierarchies. In sections \ref{largensetup} and \ref{largenresults} we describe more expensive and thus limited experiments involving a full cluster simulation to verify the small-{\it N} work. Given the history of massive star formation theory, we would like to emphasise that the processes we are studying here have very little to do directly with the formation of individual massive stars; we are studying the relatively uncontroversial dynamics of star clusters using initial conditions that can arguably be supported by the gamut of massive star formation theories. Our primary assumptions are the existence of star clusters and higher order multiples.

\section{Small-N experimental setup}
\label{smallnsetup}
For this part of the paper we set up isolated hierarchical triple systems of massive stars and bombarded them with other stars (called `intruders'). We used the code {\sc Fewbody} \citep{fregeau04}, an \nbody integrator optimised for this type of small-{\it N} experiment. While \citet{fregeau04} should be consulted for details, briefly: the code uses a variable timestep 8th order Runge-Kutta method with an embedded 7th order estimator for error control \citep{prince81}. During integration if a subsystem of the stars becomes isolated (e.g. a temporary binary is on a looping orbit far from the other stars) it is advanced analytically until the tidal force on the subsystem from other stars exceeds some fraction of the internal force of the subsystem; we use $10^{-5}$ as the tolerance for direct integration.

The hierarchical triple system consisted of an inner binary with semi-major axis $a_0$ which was in turn one member of a wider binary with $a_1 \gg a_0$. Both binaries initially had zero eccentricity, with orbital planes randomly oriented relative to each other. We chose the mass of the primary star $m_0$ from a Salpeter-like mass function $dN/dm \propto m^{-2.3}$ between 10 and 150 \msun, and the inner binary partner's mass $m_1$ by randomly selecting a binary mass ratio $q = m_1 / m_0$ in the range 0.1 to 1.0, with the mass ratio distributed as a power law with $p(q) \propto q^{-0.1}$. Recent observations of O-star binaries in six Galactic open clusters \citep{sana12b} motivated this choice. The outer binary partner's mass $m_2$ was likewise drawn by picking a mass ratio relative to the system's primary star $m_0$. The inner binary thus had mass $M_b = m_0 + m_1$, and the hierarchy as a whole had $M_t = M_b + m_2$. The intruding star's mass $M_i$ had the same mass function as the primary, but in the range 1 to 150 \msun. Table \ref{symboltable} collects these symbols for reference. 

During integrations, {\sc Fewbody} detects collisions between stars if their surfaces are detected to have touched. These collisions are mass and momentum conserving, an approach sometimes referred to as `sticky spheres'. We therefor needed a radius for each star. We used a simple parameterisation of main-sequence radii: with radii and masses here in Solar units, $r = m ^{0.9}$ for $m < 2.5$, and $r = 2.5^{0.9} (m/2.5)^{0.6}$ for $m \ge 2.5$ \citep[see e.g. figure 4 in ][]{demircan91}. We make no claim that this choice was overly realistic. Since there is no stellar evolution included in {\sc Fewbody}, we merely wanted plausible and non-extreme values. The parameterisation is probably conservatively small for very young stars that may be actively accreting \citep{hosokawa09}. 

\begin{figure*}
 \includegraphics[width=180mm]{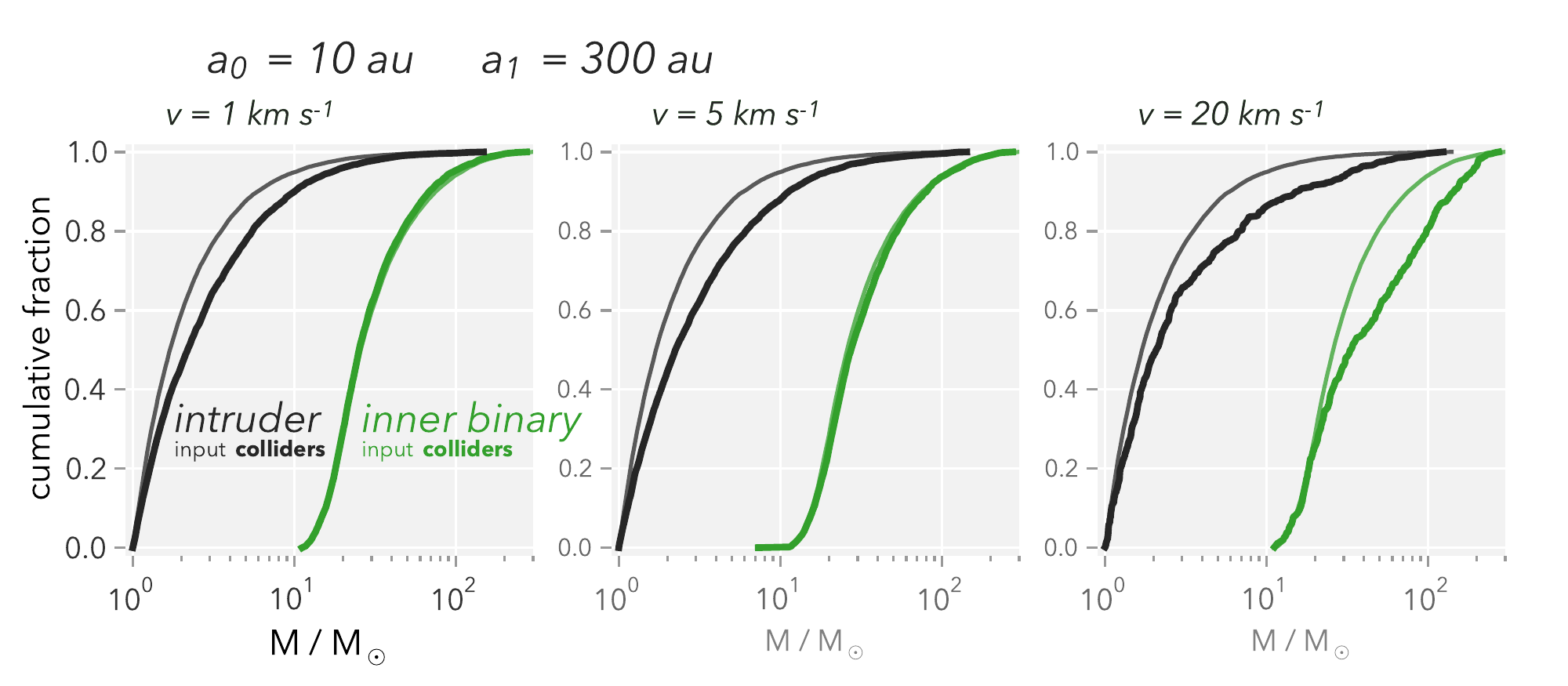}
 \caption{The intruder and inner binary mass functions for all runs (thin lines) and runs where a collision occurred (thick lines) with $a_0$ = 10 au and $a_1$ = 300 au. Collisions occur more often with more massive intruders. The mass of the hierarchy components only becomes a factor when the hierarchy is no longer energetically hard with respect to the intruder.}
  \label{masses_10_300_all}
\end{figure*}

We introduced the intruder at a random orientation relative to the hierarchy, with relative velocity at infinity $v$ and the impact parameter distributed proportionally to the probability of the encounter (i.e. proportional to the square of the impact parameter) out to a maximum value. The maximum value varied according to the stellar masses and the velocity of the encounter so that the maximum pericenter between the intruder and the hierarchy was equal to $3 a_1$. If, at the end of an encounter,  a stable triple still existed \citep[according to the approximate triple stability criterion of][]{mardling01} we sent another intruder in, randomly drawn in the same way as the first one. This process repeated until either the triple had been disrupted or else the system had survived 20 intruders. {\em Throughout the paper, if a hierarchy was reduced to singles and binaries we refer to it as `resolved'.}

For semi-major axes we used stable combinations of $a_0 / {\rm au} \in \{10, 30, 100\}$ and $a_1 / {\rm au} \in \{100, 300, 1000\}$. With $v / {\rm km~s}^{-1} \in \{1, 5, 20\}$ we covered velocities appropriate to a wide range of cluster environments. For each set of hierarchy and encounter parameters we ran 16384 random realisations, for a total of nearly $3 \times 10^5$ individual systems, each undergoing up to 20 encounters. This limit was arbitrary; the number of encounters a hierarchy may experience depends sensitively on its environment. The simulations described in section \ref{largensetup} include the appropriate encounter rate for example clusters.

\begin{table}
 \centering
 \begin{minipage}{80mm}
  \caption{Symbols describing the hierarchy and the intruder.}
  \begin{tabular}{@{}lr@{}}
  \hline
 $m_0$ & Inner binary primary mass\\
 $m_1$ & Inner binary secondary mass; $m_1 <= m_0$\\
 $m_2$ & Outer binary secondary mass; $m_2 <= m_0$\\
 $M_b$ & Inner binary mass; $M_b = m_0 + m_1$\\
 $M_t$ & Triple mass; $M_t = m_0 + m_1 + m_2$\\
 $M_i$ & Intruder mass\\
 $a_0$ & Inner binary semi-major axis\\
 $a_1$ & Outer binary semi-major axis\\
 $v$ & Relative velocity between the intruder and the triple\\
  \hline     

\end{tabular}
\label{symboltable}
\end{minipage}
\end{table}

\section{Small-N results}
\label{smallnresults}

\begin{table}
\centering
\begin{minipage}{80mm}
\caption{Collision fractions for the isolated encounter simulations.}
\begin{tabular}{@{}lllccccccc@{}}
\hline
 &  &  & \multicolumn{7}{c}{percent of runs with colliders}  \\
$a_0$& $a_1$ & v & 0:1 & 0:2 & 0:i & 1:2 & 1:i & 2:i & total \\
\hline
10 & 100 & 1 & 31.9 & 1.3 & 2.9 & 1.2 & 1.1 & 0.5 & 39.6 \vspace{0pt}\\
\multicolumn{3}{r}{91\% resolved} & 34.6 & 1.4 & 2.6 & 1.3 & 0.8 & 0.4 & 41.8 \vspace{2pt}\\
10 & 100 & 5 & 32.8 & 1.5 & 3.2 & 1.4 & 1.1 & 0.5 & 41.1 \vspace{0pt}\\
\multicolumn{3}{r}{80\% resolved} & 40.5 & 1.8 & 3.2 & 1.6 & 1.0 & 0.4 & 49.2 \vspace{2pt}\\
10 & 100 & 20 & 17.7 & 0.6 & 0.8 & 0.5 & 0.4 & 0.2 & 20.2 \vspace{0pt}\\
\multicolumn{3}{r}{42\% resolved} & 42.1 & 1.3 & 1.1 & 1.1 & 0.5 & 0.2 & 46.5 \vspace{2pt}\\
30 & 300 & 1 & 20.9 & 0.6 & 1.1 & 0.7 & 0.4 & 0.1 & 23.9 \vspace{0pt}\\
\multicolumn{3}{r}{93\% resolved} & 22.3 & 0.6 & 1.0 & 0.7 & 0.4 & 0.1 & 25.3 \vspace{2pt}\\
30 & 300 & 5 & 20.9 & 0.8 & 1.2 & 0.6 & 0.4 & 0.2 & 24.3 \vspace{0pt}\\
\multicolumn{3}{r}{70\% resolved} & 29.8 & 1.1 & 1.4 & 0.9 & 0.4 & 0.2 & 34.0 \vspace{2pt}\\
30 & 300 & 20 & 8.5 & 0.1 & 0.1 & 0.2 & 0.0 & 0.0 & 8.9 \vspace{0pt}\\
\multicolumn{3}{r}{23\% resolved} & 37.7 & 0.5 & 0.3 & 0.7 & 0.1 & 0.0 & 39.3 \vspace{2pt}\\
100 & 1000 & 1 & 11.2 & 0.3 & 0.4 & 0.3 & 0.1 & 0.1 & 12.4 \vspace{0pt}\\
\multicolumn{3}{r}{95\% resolved} & 11.8 & 0.3 & 0.4 & 0.3 & 0.1 & 0.0 & 13.0 \vspace{2pt}\\
100 & 1000 & 5 & 10.6 & 0.1 & 0.2 & 0.2 & 0.1 & 0.0 & 11.2 \vspace{0pt}\\
\multicolumn{3}{r}{48\% resolved} & 22.0 & 0.3 & 0.3 & 0.4 & 0.2 & 0.0 & 23.1 \vspace{2pt}\\
100 & 1000 & 20 & 3.5 & 0.0 & 0.0 & 0.0 & 0.0 & 0.0 & 3.5 \vspace{0pt}\\
\multicolumn{3}{r}{10\% resolved} & 33.2 & 0.2 & 0.0 & 0.2 & 0.0 & 0.0 & 33.6 \vspace{2pt}\\
\hline
10 & 300 & 1 & 23.5 & 0.7 & 2.1 & 0.5 & 0.8 & 0.4 & 28.2 \vspace{0pt}\\
\multicolumn{3}{r}{90\% resolved} & 25.6 & 0.7 & 1.7 & 0.5 & 0.6 & 0.2 & 29.7 \vspace{2pt}\\
10 & 300 & 5 & 20.4 & 0.6 & 1.5 & 0.5 & 0.6 & 0.3 & 24.0 \vspace{0pt}\\
\multicolumn{3}{r}{59\% resolved} & 34.3 & 0.9 & 1.6 & 0.7 & 0.5 & 0.2 & 38.4 \vspace{2pt}\\
10 & 300 & 20 & 2.6 & 0.0 & 0.1 & 0.0 & 0.1 & 0.0 & 2.9 \vspace{0pt}\\
\multicolumn{3}{r}{13\% resolved} & 20.1 & 0.1 & 0.2 & 0.4 & 0.1 & 0.0 & 21.1 \vspace{2pt}\\
30 & 1000 & 1 & 8.9 & 0.3 & 0.5 & 0.2 & 0.1 & 0.1 & 10.1 \vspace{0pt}\\
\multicolumn{3}{r}{95\% resolved} & 9.3 & 0.2 & 0.5 & 0.2 & 0.1 & 0.0 & 10.4 \vspace{2pt}\\
30 & 1000 & 5 & 6.0 & 0.1 & 0.3 & 0.1 & 0.1 & 0.0 & 6.6 \vspace{0pt}\\
\multicolumn{3}{r}{35\% resolved} & 17.1 & 0.2 & 0.4 & 0.3 & 0.2 & 0.0 & 18.3 \vspace{2pt}\\
30 & 1000 & 20 & 0.2 & 0.0 & 0.0 & 0.0 & 0.0 & 0.0 & 0.3 \vspace{0pt}\\
\multicolumn{3}{r}{5\% resolved} & 4.9 & 0.1 & 0.2 & 0.2 & 0.0 & 0.0 & 5.5 \vspace{2pt}\\
\hline
10 & 1000 & 1 & 5.6 & 0.2 & 0.6 & 0.2 & 0.3 & 0.1 & 7.0 \vspace{0pt}\\
\multicolumn{3}{r}{94\% resolved} & 5.8 & 0.2 & 0.5 & 0.2 & 0.3 & 0.1 & 7.1 \vspace{2pt}\\
10 & 1000 & 5 & 2.5 & 0.1 & 0.3 & 0.1 & 0.1 & 0.0 & 3.1 \vspace{0pt}\\
\multicolumn{3}{r}{28\% resolved} & 8.9 & 0.4 & 0.4 & 0.3 & 0.1 & 0.0 & 10.2 \vspace{2pt}\\
10 & 1000 & 20 & 0.1 & 0.0 & 0.0 & 0.0 & 0.0 & 0.0 & 0.1 \vspace{0pt}\\
\multicolumn{3}{r}{4\% resolved} & 2.4 & 0.3 & 0.1 & 0.1 & 0.0 & 0.0 & 3.0 \vspace{2pt}\\
\hline
\end{tabular}
Semi-major axes $a_0$ and $a_1$ are in au; velocity $v$ is in km s$^{-1}$. Collision partners `0', `1', and `2' are the primary, inner secondary, and outer secondary. Partner `i' is the impactor.
Each set shows the percentage of collisions for all runs as well as only resolved runs
\end{minipage}
\label{collisiontable}
\end{table}

Table \ref{collisiontable} summarises the collision frequency for each set of runs. For each setup we show the percent of runs with various collision partners for all of the runs as well as for only those runs in which the original hierarchy resolved to a mixture of singles and lone binaries. We regard this quantity as more fundamental than the total percentage; the overall collision percentages can only rise as more and more intruders are simulated, while the resolved fraction is representative of the overall potential for collisions. The number of intruders that might be expected in a real cluster depends on the timeframe of interest and the cluster properties. In section \ref{largensetup} we consider a limited set of these runs in a cluster setup similar to something like the Orion Nebula Cluster. We now discuss some of the trends apparent in these results; most of these are consistent with results from binary-single scattering work.

In figure \ref{masses_10_300_all} we show for one of the hierarchy setups ($a_0 = 10$ au, $a_1$ = 300 au; this is representative of all the sets) the intruder star $M_i$ and inner binary $M_b$ mass functions, showing both the input mass function and that for systems involved in collisions. At relatively low intruder velocities (v = 1 and 5 \kms) the mass function of inner binaries in colliding systems is nearly identical to the input mass function. More massive intruders are more likely to result in collisions. At high relative velocities large values of $M_b$ likewise begin to become more likely. This is due to the increased energy of the intruders. Taking the hierarchy for the moment as a binary consisting of $M_b$ and $m_2$, most systems are energetically hard to most intruders at 5 \kmss with the large hierarchy masses we are dealing with. At 20 \kmss a less massive hierarchy is not robust to ionisation, and more massive systems are needed to remain intact for several collisions.

\begin{figure}
 \includegraphics[width=80mm]{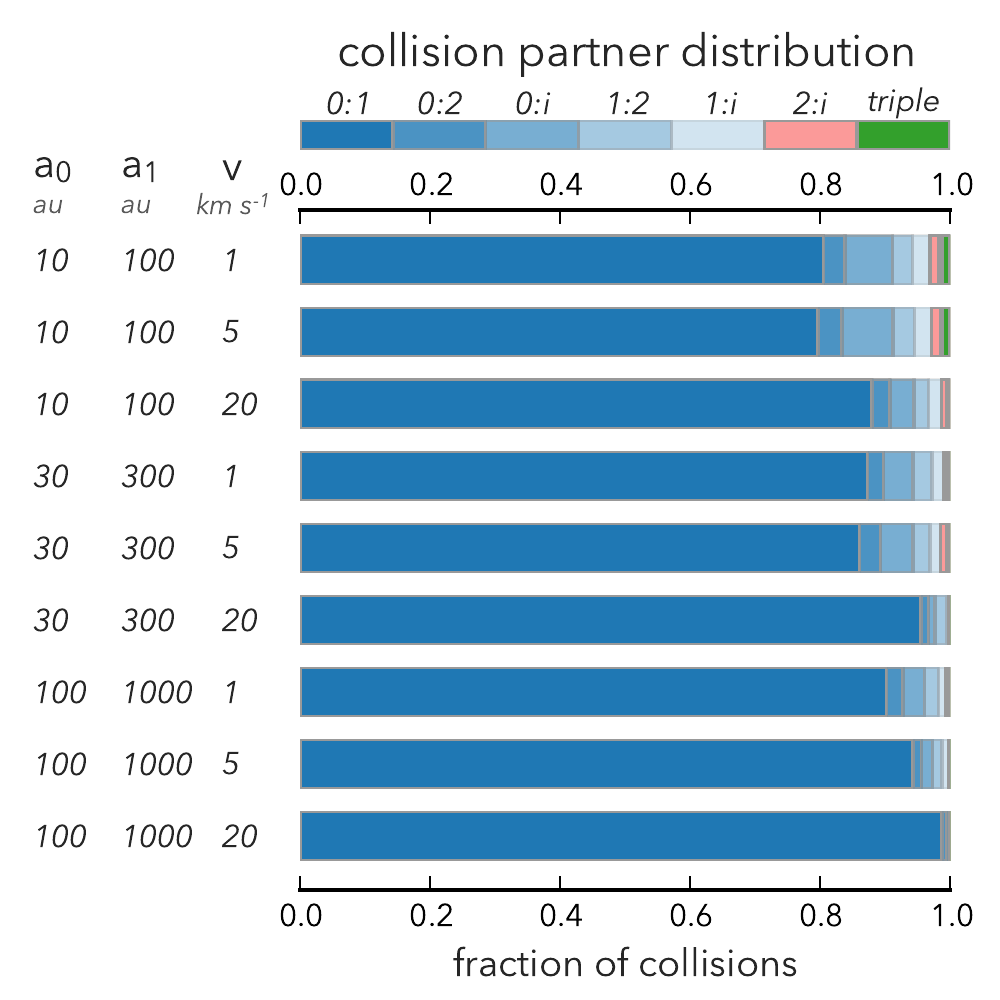}
 \caption{For those runs with collisions, the distribution of collision partners. Blue shades are double collisions involving a member of the inner binary. Red and green show the small number of collisions between the outer member of the hierarchy and the intruder and triple collisions, respectively. Collisions are predominantly between the members of the inner binary.}
  \label{collisionpartnerdistribution}
\end{figure}

When a collision does occur, the stars of the inner binary are most likely to be involved. In figure \ref{collisionpartnerdistribution} we show the fractional distribution of collision partners for the runs with $a_1 = 10 a_0$. Shades of blue are collisions involving at least one member of the inner binary, with the darkest shade indicating a collision between $m_0$ and $m_1$. Collisions involving the outer member of the hierarchy are quite rare.

\begin{figure}
 \includegraphics[width=80mm]{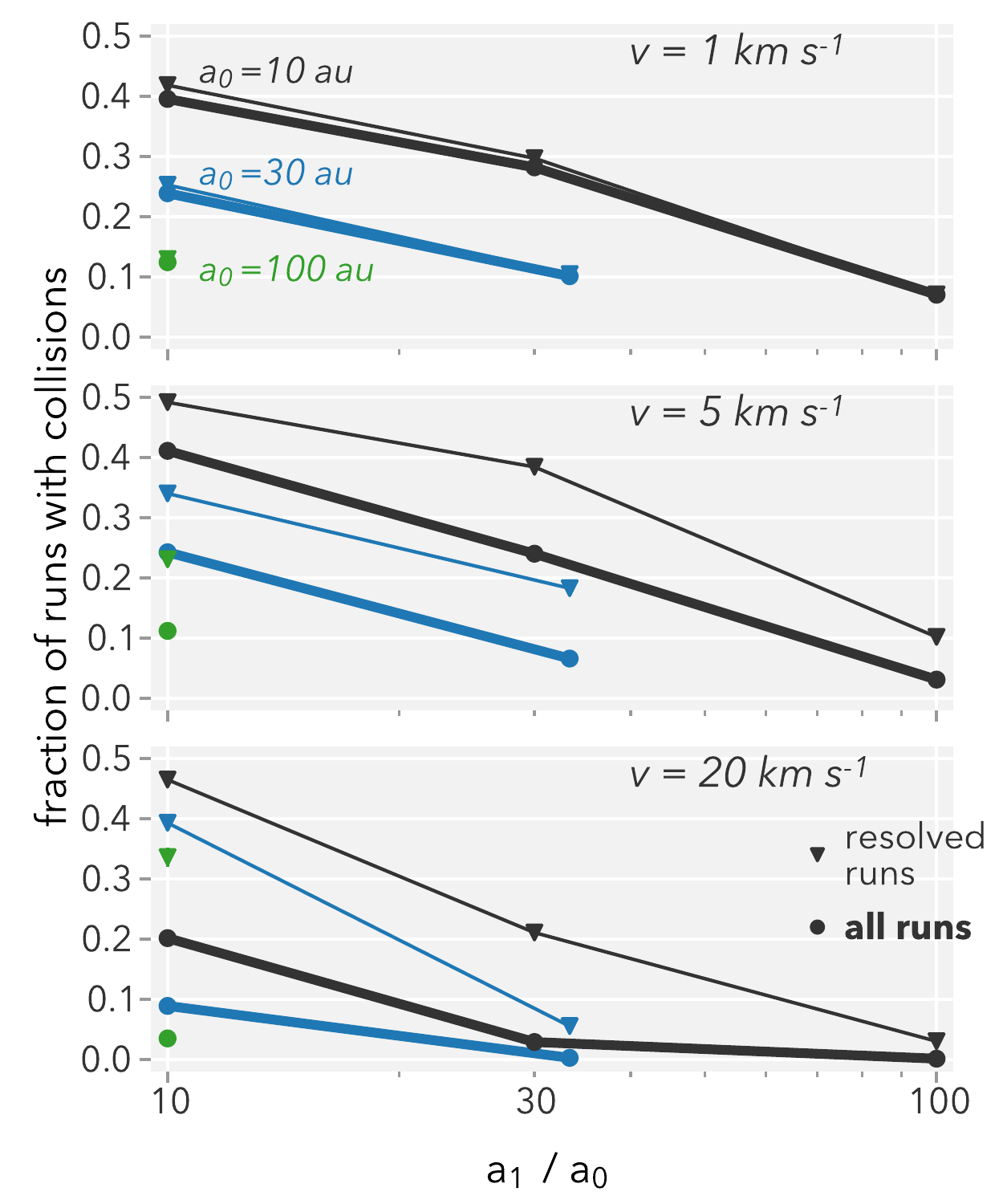}
 \caption{The fraction of runs that have a collision versus the ratio of the outer to inner binary semi-major axis. Thick lines show the result for all runs, thin lines show only resolved runs.}
  \label{semimajoraxistrends}
\end{figure}

\begin{figure}
 \includegraphics[width=80mm]{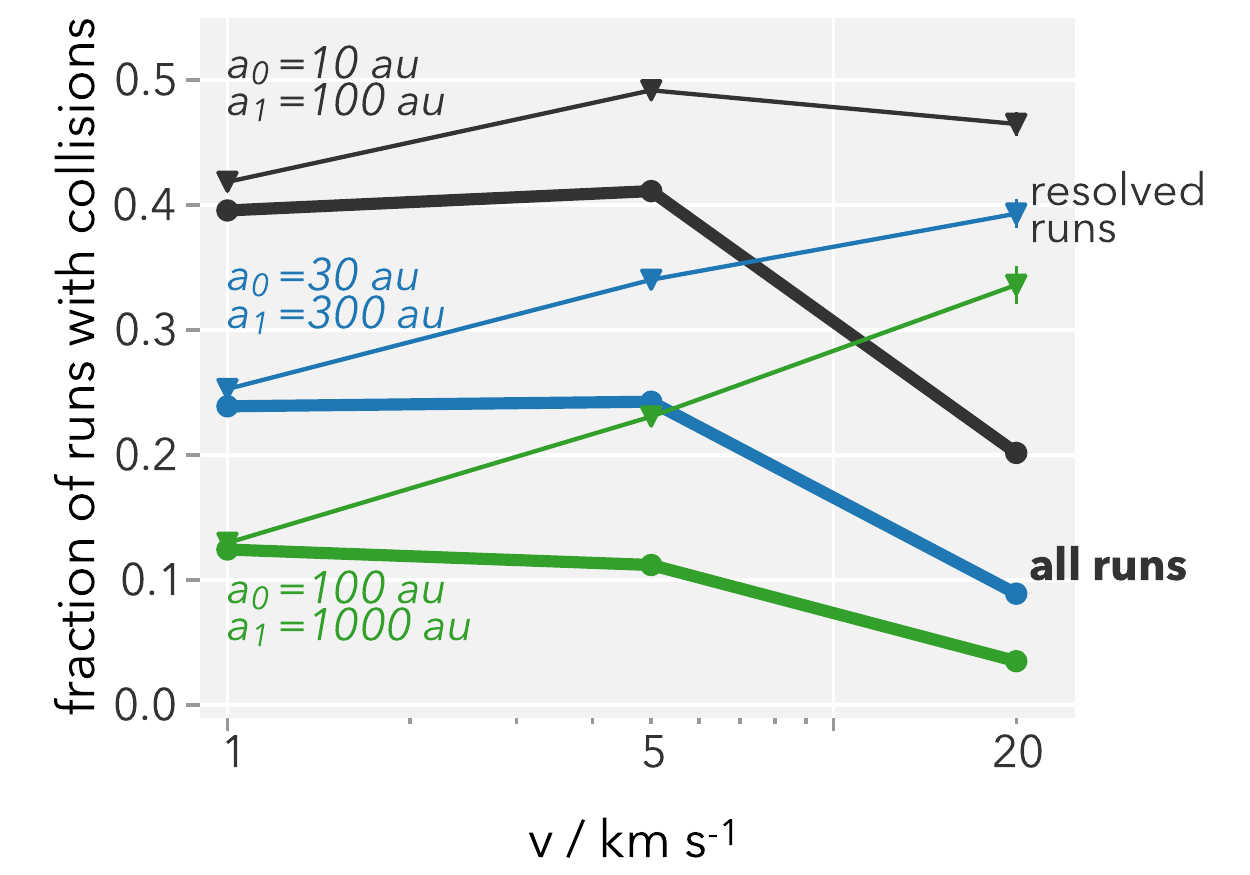}
 \caption{The fraction of runs that have a collision for all the runs with $a_1 = 10a_0$ as a function of the velocity of the intruder. Thick lines show the result for all runs, thin lines show only resolved runs.}
  \label{velocitytrends}
\end{figure}

In figure \ref{semimajoraxistrends} we plot the collision fraction as a function of the outer to inner binary ratio, $a_1 / a_0$. While we have too few points to determine the functional form of this relationship, the general trend is clear and unsurprising; the better the inner binary can be approximated as a single body the lower the collision fraction. The other relationship made clear in this plot is the decreasing collision fraction as the inner binary separation increases. This is analogous to the binary--single and binary--binary collision results of \citet{fregeau04}, who show that the collision fraction increases roughly as a weak (sub-linear but not constant) power of  $R_{\star} / a$.

Figure \ref{velocitytrends} shows the collision fractions as a function of intruder velocity, again for the runs with $a_1 = 10 a_0$. The decline in the overall fraction is, as mentioned above, mainly due to the unresolved nature of an increasing percentage of runs as velocity increases. This tendency for higher velocity runs to require more encounters to resolve the hierarchy can be explained by appeal to the existing extensive work done on binary-single scattering \citep{hut83,hut83b}. Treating the inner binary as its centre of mass particle (so that the hierarchy is reduced to a single binary with components $M_b$ and $m_2$), the critical energy at which the energy of the binary plus intruder is zero:
\begin{equation}
	v_c^2 = G \frac{M_b m_2 ( M_b + m_2 + M_i)}{M_i(M_b + m_2)}\frac{1}{a_1}.
\end{equation}
For roughly equal mass stars, this is within a factor of order unity of equation \ref{hardsoft}.
For intruder velocities $v < v_c$ encounters are likely to be more complex. As $v$ gets higher relative to $v_c$ the most likely non-flyby outcome is ionisation of the binary, but its cross section becomes increasingly small, as $v^{-2}$ \citep{hut83}. As the timescale over which each individual encounter takes place shrinks, so does the chance of close interactions between the intruder and a binary component, and more encounters are flybys that leave the binary relatively unscathed.

The decrease in the resolved fraction itself with increasing velocity makes sense in this approximation. The explanation for the increased collision fraction of resolved runs with increasing $v$ has its roots in the same idea. In order for the hierarchy to be resolved by our definition, the possible outcomes are ionisation of the outer binary from the inner binary, more complicated reorganisation of the four stars into binaries and singles, or a collision. As mentioned above the ionisation cross section decreases as $v^{-2}$, while \citet{fregeau04} showed that for binary--single and binary--binary encounters the collision cross section drops to a constant but small value consistent with the physical cross section of the individual binary components.

In this high $v$ limit, the ionisation cross section can then drop below the collision cross section, particularly in a hierarchy when one member is a binary. The problem is then effectively a binary--single scattering problem with the binary being the inner component of the hierarchy. That is, at high encounter velocities relative to the outer binary critical velocity, the increased cross section of the hierarchy no longer helps to lower the inner binary's encounter timescale. Switching focus now to the inner binary and {\em its} critical velocity, with more massive stars it will have a lower value of $v / v_c$, and thus higher collision rates \citep{fregeau04}

\begin{figure*}
 \includegraphics[width=180mm]{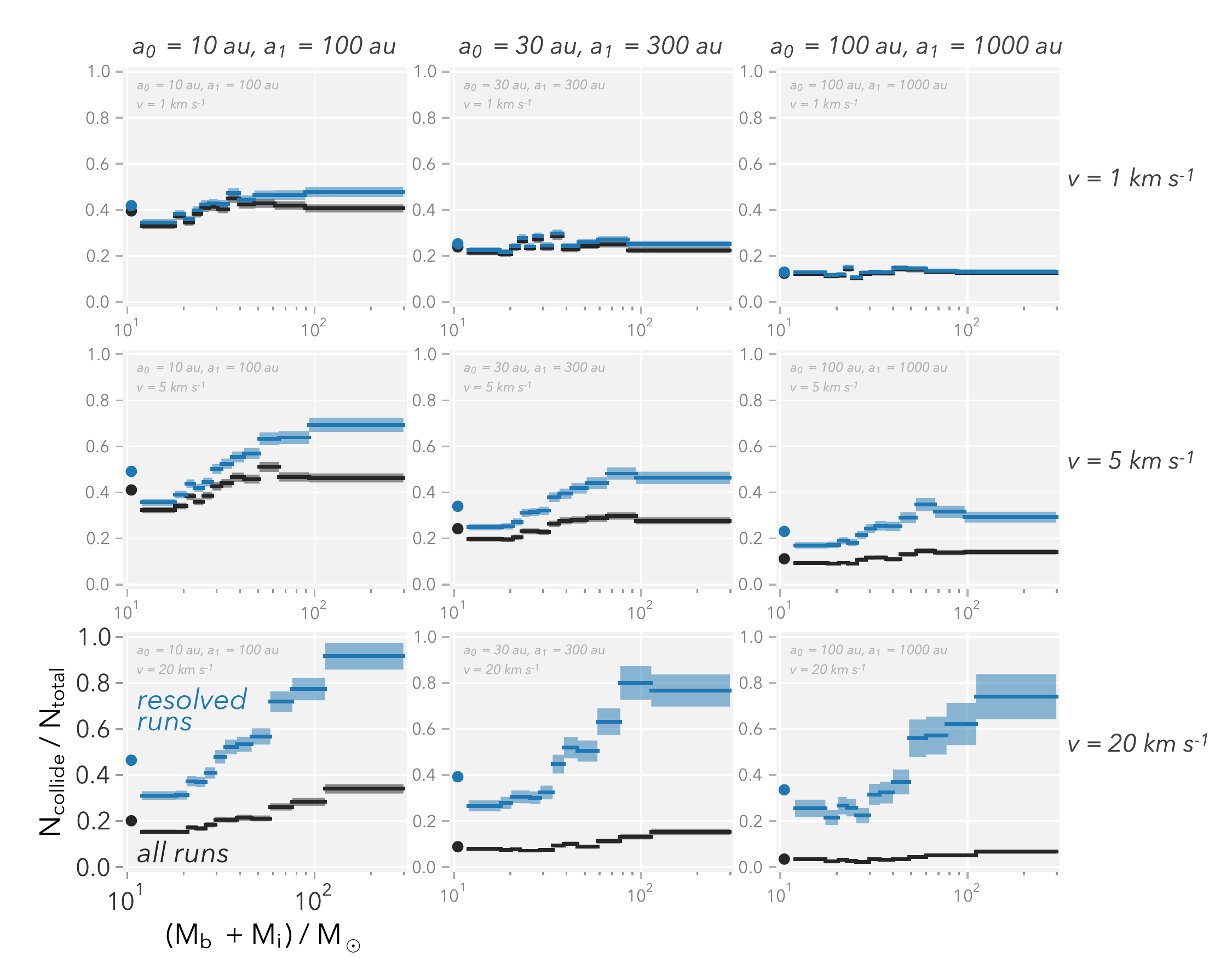}
 \caption{The fraction of runs with collisions versus the mass of the inner binary plus intruder for all the runs with $a_1 = 10a_0$. The shaded regions show the uncertainty in the estimate assuming Poisson counting statistics. The fractions are shown for all the runs done (black lines) and only resolved runs with no remaining triple systems (blue lines). Mean values over all bins are shown as the points at 10 \msun. }
  \label{collisionfractiongrid}
\end{figure*}

This is seen in figure \ref{collisionfractiongrid}, showing the fraction of colliding systems as a function of $M_b + M_i$ for the $a_1 = 10 a_0$ runs for all three velocities (we plot these masses rather than the system total since the outer hierarchy member seldom took part in a collision). As $v$ increases, the fraction of resolved runs drops, but those that {\em are} resolved are more likely to be resolved due to collisions than by disruption of the outer binary. The most massive systems have lower encounter velocities relative to the inner binary $v_c$, and thus a higher collision fraction. For example, an inner 30 au binary with 20 \msuns components encountering a 5 \msuns impactor at $v = 20$ \kms has $v/v_c \sim 0.4$. If we change the binary masses to 2 \msun, and $v/v_c \sim 2.8$.  As $v$ increases the problem reduces to an encounter with a single binary (the inner one), and the resolved collision fraction for different hierarchies start to converge, seen in figure \ref{velocitytrends}.

These arguments are clearly approximate, treating the hierarchy as disconnected binaries, but the results seem to fall in line with the extensive work done on cleaner binary--single and binary--binary encounters. The details of triple--single encounters have only recently begun to be explored in the same way \citep{leigh12}, and more idealised experiments than those we have performed here (e.g. a more controlled mass function) are needed to tease out the fundamental details. The more naturalistic setup we have chosen gives some hints about the trends that might be expected for the specific case of massive hierarchies in a young cluster. The question then becomes, how many hierarchies will be resolved in a realistic cluster, rather then the arbitrary 20 encounters that we have performed?

\section{The coplanar special case}
\label{coplanarscattering}
The work we discussed above used hierarchies with randomly oriented orbital planes. While this is the most general case and appropriate to dynamically formed hierarchies, in a disc formation scenario nearly coplanar orbits might be expected. This special case caries with it two important physical effects. First, coplanar orbits maximise the net angular momentum of a hierarchy. Previous scattering work has shown that the angular momentum of a system is a key parameter in determining the scattering outcome \citep{mikkola86}. Specific to the situation at hand, \citet{leigh12} showed that in triple--single encounters increasing the total angular momentum decreases the collision probability.

Second, coplanar orbits suppress the Kozai-Lidov mechanism \citep{kozai62,lidov62,naoz11} in which secular perturbations can lead to extreme oscillations in the inner eccentricity of a inclined hierarchical systems like we study here. The timescale over which the oscillations take place is of order \citep{kiseleva98}
\begin{equation}
	\tau_{KL} = \frac{2 P_1^2}{3 \pi P_0^2} (1-e_1^2) \frac{m_0 + m_1 + m+2}{m_2},
\end{equation}
with $P_0$ and $e_0$ the period and eccentricity of the inner binary, and subscript 1 the outer binary.
For the system masses and sizes we used, $\tau_{KL}$ can be of order $10^5$ yr. In general, the cumulative integration time of the small-{\it N} experiments was shorter than this, and the mechanism was already somewhat suppressed. Over the 2 Myr large-{\it N} runs that we discuss below, some of the systems would be able to go through one or more Kozai cycles, and potentially become tidally interacting or merge. 

\begin{table}
\centering
\begin{minipage}{80mm}
\caption{Collision fractions for the {\em coplanar} isolated encounter simulations.}
\begin{tabular}{@{}lllccccccc@{}}
\hline
 &  &  & \multicolumn{7}{c}{percent of runs with colliders}  \\
$a_0$& $a_1$ & v & 0:1 & 0:2 & 0:i & 1:2 & 1:i & 2:i & total \\
\hline
10 & 100 & 3 & 11.7 & 1.1 & 3.7 & 1.4 & 1.5 & 0.7 & 20.4 \vspace{0pt}\\
\multicolumn{3}{r}{80\% resolved} & 14.5 & 1.3 & 3.2 & 1.7 & 1.3 & 0.5 & 22.9 \vspace{2pt}\\
10 & 300 & 3 & 8.2 & 0.6 & 2.2 & 0.5 & 1.0 & 0.5 & 13.1 \vspace{0pt}\\
\multicolumn{3}{r}{69\% resolved} & 11.7 & 0.7 & 1.8 & 0.6 & 0.6 & 0.4 & 16.0 \vspace{2pt}\\
10 & 1000 & 3 & 2.8 & 0.2 & 0.6 & 0.2 & 0.3 & 0.1 & 4.2 \vspace{0pt}\\
\multicolumn{3}{r}{56\% resolved} & 4.8 & 0.1 & 0.5 & 0.3 & 0.2 & 0.0 & 6.0 \vspace{2pt}\\
30 & 300 & 3 & 6.8 & 0.5 & 1.5 & 0.9 & 0.6 & 0.2 & 10.5 \vspace{0pt}\\
\multicolumn{3}{r}{77\% resolved} & 8.7 & 0.5 & 1.4 & 1.1 & 0.6 & 0.2 & 12.6 \vspace{2pt}\\
30 & 1000 & 3 & 2.6 & 0.1 & 0.7 & 0.2 & 0.2 & 0.1 & 4.0 \vspace{0pt}\\
\multicolumn{3}{r}{59\% resolved} & 4.1 & 0.2 & 0.8 & 0.3 & 0.2 & 0.1 & 5.9 \vspace{2pt}\\
\hline
\end{tabular}
Semi-major axes $a_0$ and $a_1$ are in au; velocity $v$ is in km s$^{-1}$. Collision partners `0', `1', and `2' are the primary, inner secondary, and outer secondary. Partner `i' is the impactor.
Each set shows the percentage of collisions for all runs as well as only resolved runs
\end{minipage}
\label{collisiontablecoplanar}
\end{table}

As a topic of future work, the interplay between tidal dissipation and the secular dynamics of hierarchical systems may be interesting in the case of massive primordial triples. The Kozai cycle with tidal friction (KCTF) mechanism has been extensively studied in lower-mass stars and planetary systems \citep[e.g.][]{mazeh79,kiseleva98,eggleton01,fabrycky07,perets09}. The short Kozai periods (relative to the main sequence lifetime of a massive star) and large radii of massive stars suggests that the process could be important for producing massive short period binaries. This is not the point of our work here, however. In order to remove the possibility of merging inner binaries due solely to the internal dynamics of the hierarchy, we used co-planar hierarchies in the cluster experiments to remove the inclination--eccentricity exchange. 

In order to make a direct comparison between the cluster runs and the scattering results, we ran a limited set of scattering experiments directly tailored to the cluster characteristics that we describe fully below. While otherwise identical to the previous scattering results, these ones used coplanar hierarchies with $v = 3$. We performed 8096 runs each of the following semi-major axis pairs: $[a_0, a_1] = [30, 300], [30, 1000], [10, 100], [10, 300]$ and $[10, 1000]$ au. Table \ref{collisiontablecoplanar} shows the results of these runs; the collision rate is about one half that of the general case, and the general trends with semi-major axes are similar.

\section{Large-N experimental setup}
\label{largensetup}
The small-{\it N} experiments we ran were necessarily highly idealised (a clustered environment was approximated by repeatedly sending in single intruders). This was necessary in order to run a large number of experiments in a reasonable amount of time, and desirable in order to control the parameters of the encounters. In order to verify the results in a more `realistic' setting, we ran a limited set of full cluster experiments using the \nbody code {\sc nbody6} \citep{aarseth00, aarseth03}. The code uses the 4th order Hermite algorithm \citep{makino92} to integrate the stars, augmented by algorithmic regularisation of close encounters and binaries \citep{kustaanheimo65}. Stellar evolution is included via lookup tables \citep{hurley00}, providing the radii of the stars used in collision detection. 

\begin{figure}
 \includegraphics[width=80mm]{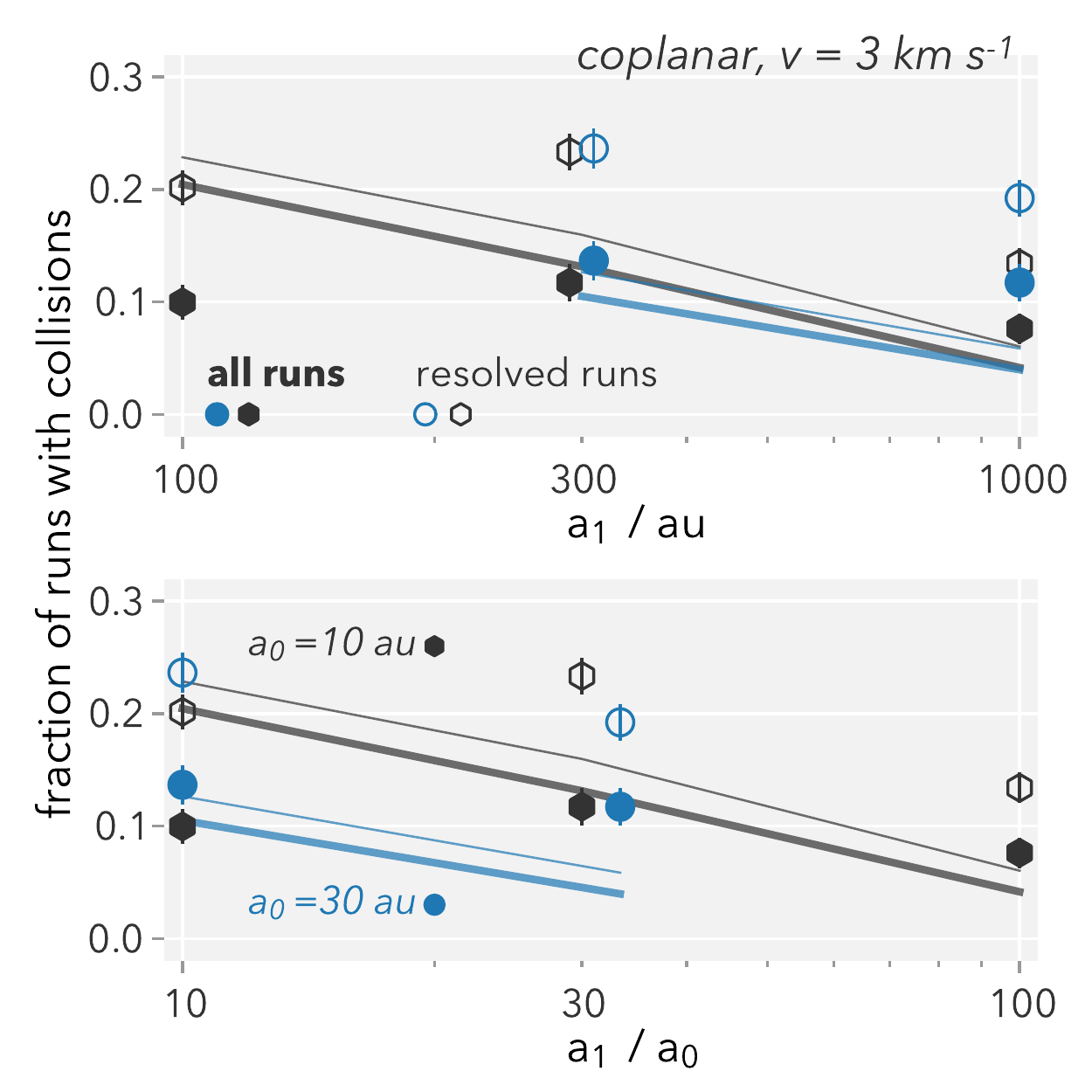}
 \caption{Symbols show the fraction of cluster runs that have a collision versus the outer semi-major axis $a_1$ (top), and the ratio of the hierarchy's outer to inner binary semi-major axis (bottom), plotted over the coplanar small-{\it N} scattering results shown as lines. The results at $a_1 = 300$ au have been offset horizontally for clarity.}
  \label{semimajoraxistrends_cluster}
\end{figure}

We set up clusters of 2048 stars using the mass function described in \citet{maschberger12} between 0.1 and 150 \msun. This mass function is effectively the same as e.g. a \citet{kroupa01} or \citet{chabrier03} mass function; we chose it mainly for computational ease. The mean mass of this mass function is $\sim 0.65$ \msun, giving a cluster mass of $~\sim 1300$ \msun, although the exact value was not fixed. We used \citet{king66} models with $W_0 = 6$ and a central mass density of $2 \times 10^4$ \msuns pc$^{-3}$. This translated to half-mass radii of $\sim$ 0.4 pc, and one dimensional velocity dispersions of $\sim 1.5$ \kms. These characteristics were chosen to be broadly match the core properties of a young massive star forming region such as the Orion Nebula Cluster \citep{hillenbrand98}. Since the encounter velocities should have been expected to be distributed roughly as a Maxwellian with dispersion $\sqrt{2} \sigma_1 \approx 2$ \kms and mode $\sim 3$ \kms, these results should be directly comparable to the $v = 3$ coplanar small-{\it N} results. 

We initialised the hierarchical systems in the following manner: after randomly selecting a star from those that were more massive than 10 \msuns and inside roughly the half mass radius, we chose a binary mass ratio from the same power law as the small-{\it N} experiments, $p(q) \propto q^{-0.1}$. We then adjusted the mass of the star that most closely matched this target mass to match it exactly, and moved it into orbit around the primary. We chose the third star's mass in the same way and placed it in a circular coplanar orbit around the inner binary's centre of mass. The hierarchy separations we used were the same as the coplanar scattering results: $[a_0, a_1] = [30, 300], [30,1000], [10, 100], [10, 300]$ and $[10,1000]$ au. 

We ran 512 of these cluster simulations for each set of hierarchy separations, each for 2 Myr, and checked for collisions involving the hierarchy members. As controls we also ran 512 realisations each of identical setups but with only a 10, 30, 300 or 1000 au binary included.

\section{Large-N results}
\label{largenresults}
As for the small-{\it N} experiments, for each run we checked for collisions involving members of the original hierarchy. If no collisions occurred we also checked if the hierarchical system was still intact by searching for hierarchies including at least one of the original inner binary members. This allowed us to determine both a collision fraction for the specific cluster setup that we simulated as well as an approximation to the `resolved only' results presented earlier. Table \ref{clustertable} shows the results for both the hierarchy runs and the control runs. In figure \ref{semimajoraxistrends_cluster} we plot the collision fractions on top of the coplanar $v=3$ \kmss scattering results. The filled symbols show the fraction of collisions for all the cluster simulations, and the open symbols are the results for just the resolved runs.

Across the full range of hierarchies we simulated, both the total and resolved collision fractions in the cluster runs are remarkably constant, with little evidence of the trends with semi-major axes seen in the scattering results. While there may be a hint of a lower collision fraction among resolved runs as $a_1$ increases from 300 to 1000 au, the $a_0 = 10$, $a_1 = 100$ au runs had fewer collisions than their $a_1 = 300$ au counterparts.
We suspect the reason for this disagreement lies in the breakdown of the small-{\it N} assumptions in the (simulated) reality of a cluster environment. With interaction radii of order 1000 au, the mean interstellar separation in the cores of the clusters were only a factor of a few larger than the encounter radius. The extremely short interaction times there increased the number of stars that could be taking part in an interaction, and the encounters were no longer well approximated as a four-body interaction. In the top panel of figure \ref{semimajoraxistrends_cluster}, note that the relative disagreement between the scattering results and the cluster results increases with $a_1$, with only the smallest hierarchy ($a_1 = 100$ au) displaying agreement between the clusters and the isolated encounters.

This situation is reminiscent of the work of \citet{tanikawa12}, who showed that the actual formation of hard binaries in a core collapsed cluster involves many stars and bears little resemblance to the three-body approximation that is frequently used. Furthermore, with the very massive systems and low velocities we considered gravitational focussing is dominant, and the impact parameters used in the scattering experiments were of the order of the size of the cluster core. It is not clear that the encounter geometries and characteristics in the cluster were well-described by the idealised scattering runs.

For all of the setups, just over one-half of the hierarchies resolved down to binaries and singles over the 2 Myr integration. 
If we had increased the integration time of the runs we expect that the collision fraction would have converge to the resolved-only result. In practice stellar evolution begins to complicate the interpretation of the runs past a few Myr, one of the reasons for our choice of a 2 Myr cutoff. The overall collision rates for all but the $a_0 = 10$, $a_1 = 1000$ au runs were roughly 2--3 times those of the binary-only runs. For the clusters (and integration times) that we simulated, the enhancement to collisions caused by the hierarchies was significant, but was a factor of order unity rather than an order of magnitude.

\begin{figure}
 \includegraphics[width=80mm]{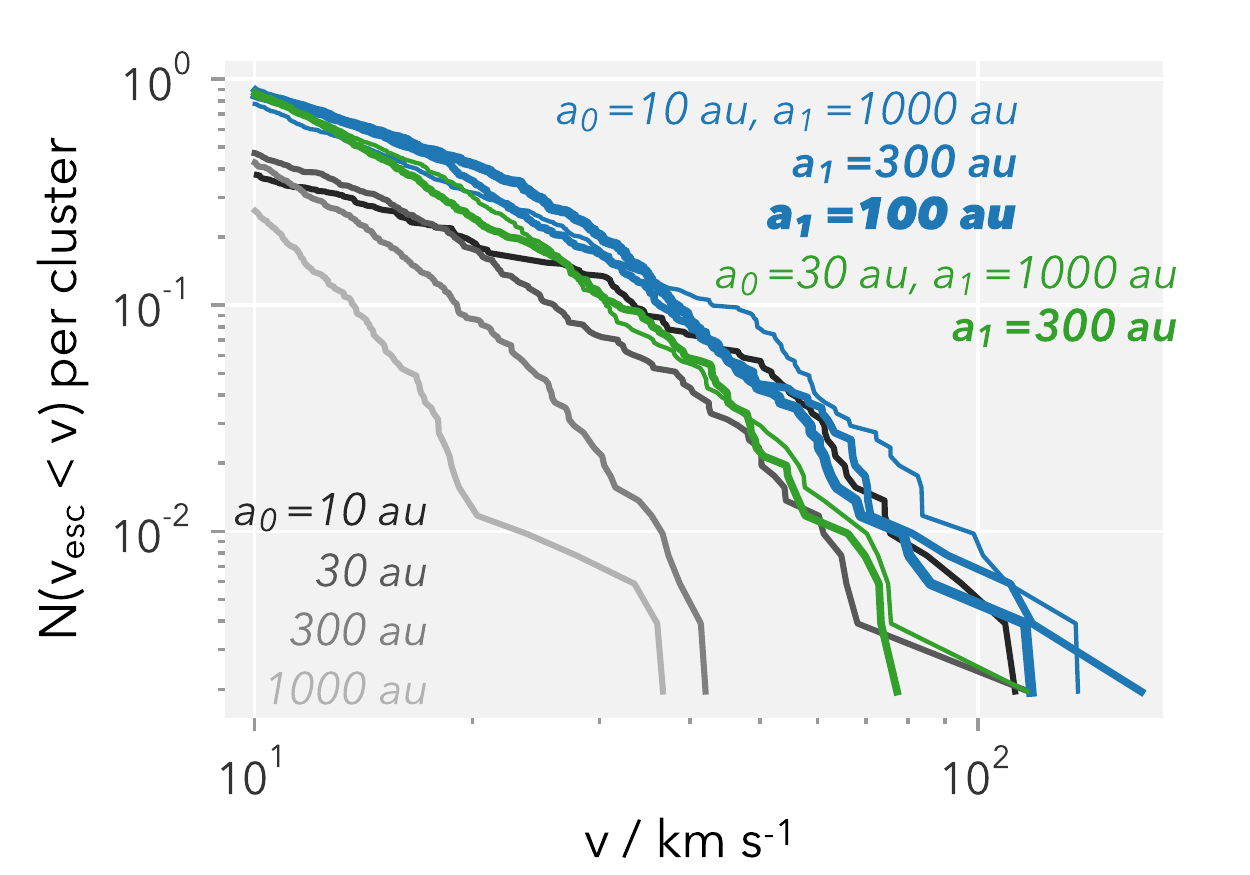}
 \caption{The mean number of stars per cluster with mass greater than 1 \msuns that escape velocity $v_{esc}$ greater than some value. Runs with initial binaries only are the lower curves in shades of gray.}
  \label{escapevelocities}
\end{figure}

Besides collisions, the heightened interaction that the inner binaries experience relative to their non-hierarchical counterparts should result in a higher frequency of stars escaping the cluster. In figure \ref{escapevelocities} we plot the number of stars ejected per cluster with velocity greater than a given value, for escapers with mass greater than 1 \msuns only. Our escaper selection was simple; stars at radii greater than 5 initial half mass radii and velocities $v_{esc} > 10$ \kmss were chosen (10 \kmss is about 5 times the escape velocity at 5 half mass radii, and safely outside the bulk of the cluster's velocity distribution; a star at that velocity would be unambiguously identified as a high velocity escaper). As expected the maximum escape velocity increases with the binding energy of the binary for the binary-only runs, and with the binding energy of the inner binary for the hierarchies. Notably, the clusters hosting hierarchical systems eject high velocity escapers at a rate 2 -- 3 times greater than the binary-only clusters.

While this result is limited (a single cluster configuration hosting a single hierarchy with two different architectures), the issue may be worthy of more detailed study elsewhere. It suggests that detailed examinations of escaper quantities may be more sensitive to the initial conditions of massive stellar systems than has been appreciated. This issue is beginning to become more important as models of individual clusters grow more sophisticated, and variations in these quantities by factors of order unity are used to discriminate between different formation scenarios \citep[e.g][]{fujii12a}.

\begin{table}
\centering
\begin{minipage}{80mm}
\caption{Collision fractions for the full cluster simulations.}
\begin{tabular}{@{}lllccccccc@{}}
\hline
 &  &  \multicolumn{6}{c}{percent of runs with colliders}  \\
$a_0$& $a_1$ & 0:1 & 0:2 & 0:i & 1:2 & 1:i & 2:i & total \\
\hline
10 & 100 & 3.7&0.6&0.4&3.3&1.0&1.0&10.0\vspace{0pt}\\
\multicolumn{2}{r}{49\% resolved}&7.5&1.2&0.8&6.7&2.0&2.0&20.2\vspace{2pt}\\
10 & 300 & 6.4&0.8&0.0&2.7&0.8&1.0&11.7\vspace{0pt}\\
\multicolumn{2}{r}{50\% resolved}&12.8&1.6&0.0&5.4&1.6&1.9&23.3\vspace{2pt}\\
10 & 1000 & 4.1&0.2&0.2&1.8&1.0&0.4&7.6\vspace{0pt}\\
\multicolumn{2}{r}{57\% resolved}&7.2&0.3&0.3&3.1&1.7&0.7&13.4\vspace{2pt}\\
30 & 300 & 7.0&1.2&1.4&3.5&0.2&0.4&13.7\vspace{0pt}\\
\multicolumn{2}{r}{58\% resolved}&12.2&2.0&2.4&6.1&0.3&0.7&23.6\vspace{2pt}\\
30 & 1000 & 5.7&1.2&0.0&2.0&2.5&0.4&11.7\vspace{0pt}\\
\multicolumn{2}{r}{61\% resolved}&9.3&1.9&0.0&3.2&4.2&0.6&19.2\vspace{2pt}\\
10 & none & 2.3&--&2.1&--&0.4&--&4.9\vspace{2pt}\\
30 & none & 1.4&--&1.8&--&1.0&--&4.1\vspace{2pt}\\
300 & none & 1.4&--&2.9&--&0.6&--&4.9\vspace{2pt}\\
1000 & none & 0.8&--&1.8&--&0.4&--&2.9\vspace{0pt}\\
\hline

\end{tabular}
Semi-major axes $a_0$ and $a_1$ are in au; velocity $v$ is in km s$^{-1}$. Collision partners `0', `1', and `2' are the primary, inner secondary, and outer secondary. Partner `i' is the impactor.
Each set shows the percentage of collisions for all runs as well as only resolved runs
\label{clustertable}
\end{minipage}
\end{table}

\section{Conclusions}
After their formation wide, massive binaries, in typical young open clusters with modest velocity dispersions, will survive many encounters during their main sequence lifetime. If the binary is instead a hierarchical triple, collisions may occur with interestingly high frequency over short timescales of only  a few Myr. We carried out two series of numerical experiments to determine the collision rates of massive hierarchical star systems in young clusters.

Our small-{\it N} scattering experiments suggest that if a hierarchical system is continuously bombarded by intruding stars until it resolves into binaries and singles, then tens of percent of the hierarchies will experience a collision. The original spacing of the hierarchy and the velocity of the intruders determine the precise value. The question is then, how many hierarchies will  get resolved into binaries and singles in a typical cluster environment? In our large-{\it N} simulations of a cluster similar to something like the Orion Nebula Cluster, roughly 50 per cent of our initial hierarchical systems were broken up into binaries and singles in 2 Myr, with an overall collision fraction between about 7 and 14 per cent depending on the initial hierarchy configuration. For the specific cluster we simulated, these numbers represent a modest increase to the collision fraction for lone binaries, which were under 5 per cent. The overall agreement between the scattering results and the full cluster simulations is at best suggestive. We suspect this is due to the breakdown of the scattering experiments' isolated-encounter assumption when the hierarchies are placed in a dense cluster; a lower-density environment simulated over longer time periods might provide a better match, but then the issue of the massive stars' short lifetimes rears its head.

In addition to collisions, the high rate of encounters produces a similar excess of runaway stars compared to systems with no primordial hierarchies. The relative scarcity of collision products and high velocity escapers produced by a cluster can make them useful as a diagnostic of different cluster initial conditions or formation scenarios. Our results suggest that the higher order multiplicity characteristics of massive stars are an important input to any dynamical experiments in which they play a role in producing these rare events.

\section*{Acknowledgments}
We are grateful to Sverre Aarseth and John Fregeau for developing and making available the \nbody tools used in this work, and to Nathan Leigh and Aaron Geller for providing a copy of their triple--single scattering routine. Further thanks to Nathan Leigh and Cathie Clarke for helpful comments on the draft. We began this work at the International Space Science Institute in Bern, during a meeting of Team Goodwin; our thanks to them for providing a comfortable and productive place for collaboration.

\bsp

\label{lastpage}

\end{document}